\documentclass[aps,pra,letterpaper,superscriptaddress,twocolumn,floatfix]{revtex4-2}
\usepackage{graphicx}
\usepackage{dcolumn}

\usepackage{bm}
\usepackage{makecell}

\usepackage{subfigure}
\usepackage{xcolor}
\newcommand{\be}{\begin{eqnarray}}
\newcommand{\ee}{\end{eqnarray}}
\usepackage{amsfonts}
\usepackage{bbm}
\usepackage{amsmath,leftidx}
\usepackage{graphicx}
\usepackage{times}
\usepackage{CJK}
\usepackage{color,colortbl}
\usepackage[normalem]{ulem}
\usepackage{hyperref}

\graphicspath{ {./Figures/} }

\begin{document}

\title{
Excitation spectrum of  spin-1 Kitaev spin liquids 
}

\author{ Yu-Hsueh Chen}
\affiliation{ Department of Physics and Center for Theoretical Physics, National Taiwan University, Taipei 10607, Taiwan }
\author{ Jozef Genzor}
\affiliation{ Department of Physics and Center for Theoretical Physics, National Taiwan University, Taipei 10607, Taiwan }
\author{Yong Baek Kim}
\affiliation{Department of Physics, University of Toronto, Toronto, Ontario M5S 1A7, Canada}
\author{ Ying-Jer Kao}
\affiliation{ Department of Physics and Center for Theoretical Physics, National Taiwan University, Taipei 10607, Taiwan }
\affiliation{Physics Division, National Center for Theoretical Science, National Taiwan University, Taipei 10607, Taiwan }
\begin{abstract}
We study the excitation spectrum of the  spin-1 Kitaev model using the symmetric tensor network. 
By evaluating the virtual order parameters defined on the virtual Hilbert space in the tensor network formalism,  we confirm the ground state is in a $\mathbb{Z}_2$ spin liquid phase.
Using the correspondence between the transfer matrix spectrum and low-lying excitations,  we find  that contrary to the dispersive Majorana excitation in the spin-1/2 case, the isotropic spin-1 Kitaev model has  a dispersive charge anyon excitation.
Bottom of the gapped  single-particle charge excitations are found at $\mathbf{K}, \mathbf{K}'=(\pm2\pi/3, \mp 2\pi/3)$, with a corresponding correlation length of $\xi \approx 6.7$ unit cells.
The lower edge of the two-particle continuum, which is closely related to the dynamical structure factor measured in inelastic neutron scattering experiments, is obtained by extracting the excitations in the vacuum superselection sector in the anyon theory language.
\end{abstract}

\maketitle



Quantum spin liquids (QSLs) are phases of matter characterized by the existence of long-range entanglement in the ground states and the fractionalized excitations \citep{Savary_2016,RevModPhys.89.025003,Broholmeaay0668}. 
The exactly solvable model introduced by Kitaev\citep{Kitaev_2006} opens up a new avenue to search for QSLs materials in nature that realize Kitaev-like interactions. 
Guided by the microscopic mechanism with  strong spin-orbit coupling and $S = 1/2$ local  moments to generate bond-dependent Ising interactions \cite{PhysRevLett.102.017205}, several candidate materials have been proposed
\citep{PhysRevLett.108.127203, doi:10.1146/annurev-conmatphys-020911-125138,doi:10.1146/annurev-conmatphys-031115-011319, Winter_2017, PhysRevB.90.041112}. 
In particular, the quantized plateau in the thermal Hall conductivity observed in  $\alpha$-\text{RuCl}$_3$ \citep{Kasahara_2018} indicates the existence of the Majorana edge modes,  and shows a strong evidence of a Kitaev-like spin liquid in this material.
On the other hand, higher-spin Kitaev model has also been theoretically studied, as the frustrated Ising-like interactions may provide an alternative route to access QSLs \citep{Baskaran_2008,PhysRevB.99.104408, PhysRevB.98.214404}. 
Recently, microscopic mechanism to realize a $S = 1$ Kitaev model and  candidate materials have been proposed \citep{PhysRevLett.123.037203}, raising the importance of the study of the higher-spin Kitaev physics. 
Different from its spin-$1/2$ counterpart, the higher-spin Kitaev model cannot be exactly solved by mapping the spins to Majorana fermions, and numerical studies have been carried out to identify the nature of the  ground states for the $S = 1$ Kitaev model \citep{2018_exact_spin1,PhysRevB.102.121102,PhysRevResearch.2.022047,PhysRevResearch.3.013160, 2020-spin-one-kitaev,lee2020anisotropy}. 
While several studies suggest that the isotropic spin-1 Kiatev model exhibits spin liquids with a $\mathbb{Z}_2$ gauge structure, quantitative features about the fractionalized excitations, i.e., the excitations spectrum, are still missing.%

The excitation spectrum is deeply connected to the experiments.
If the system harbors fractionalized excitations, the dynamical spin structure factor measured in inelastic neutron scattering (INS) should exhibit a broad continuum arising from multi-particle excitations.
For example, in the case of CsCuCl$_4$, the INS experiment gives the tell-tale signature of the $S=1/2$ spinons from the dispersive continuum of excitations~\cite{Coldea:2001qv}.
In the two-dimensional (2D) system, gapped excitations of the QSLs are called anyons, and different types of anyons are distinguished by different superselection sectors \citep{Kitaev_2006,kitaev2009topological}.
To be specific, two particles are in the same sector if there exists a local operator which can transform from one to another.
Take the spin-$1/2$ Kitaev model as an example, where the quasiparticles are $\mathbb{Z}_2$ vortices and fermions \citep{Kitaev_2006}.
In the anisotropic limit, the system lies in a $\mathbb{Z}_2$ QSL phase with four superselection sectors: vacuum$(I)$, charge$(e)$, flux$(m)$, and fermion$(\epsilon)$. 
In particular, while charge and flux anyons both correspond to the $\mathbb{Z}_2$ vortices (but live in alternating rows of hexagons), they belong to different sectors since there exist no local operator to transform from one to another.
%
%
Along this line, one can also deduce that all the even-particle charge, flux, and fermion excitations belong to the same vacuum sector due to the fusion rules $e\times e = m \times m = \epsilon \times \epsilon = I$.

While there is no local order parameter to characterize QSLs, it turns out that the projected entangled pair states (PEPS)~\citep{Verstraete2004RenormalizationAF}, a type of tensor network (TN), can encode  topological properties into the symmetries of a local tensor's virtual Hilbert space.  \citep{2011_Norbert_Ginjective, Bultinck_2017}. 
To be specific, given a local tensor $A^i_{\alpha \beta \gamma \delta}$ with $i$ the physical indices and $\alpha, \beta, \gamma, \delta$ the virtual indices, the topological properties of a translational invariant PEPS $\left|\psi_A\right\rangle=\sum_{i_1,i_2,\ldots} \operatorname{tTr}\left(A^{i_1}A^{i_2}\ldots \right)|i_1, i_2,\ldots\rangle,$ (here the tensorial trace is over the virtual indices) depend on the actions purely on the virtual legs that make $A$ invariant. 
One can even define the \textit{virtual order parameters} within the PEPS framework to identify the nature of QSLs \citep{2017_anyon_condensates, 2020_order_parameter}.
On the other hand, it was found that the transfer matrix (TM), a central object measuring ground state's correlations defined in TN, contains signatures of low-lying excitations \citep{2015_Zauner}. 
This property, combined with the virtual symmetries of the PEPS tensor, can be used to extract the anyonic excitation spectrum of different superselection sectors \citep{Haegeman_2015}.

In this \textit{letter}, we compute the excitation spectrum for the isotropic spin-1 Kitaev model, exploiting the correspondence between the TM spectrum and low-energy excitations developed in the TN formalism. 
We construct the $\mathbb{Z}_2$-invariant PEPS \citep{2011_Norbert_Ginjective,Schuch_2012} to represent the spin-$1$ Kitaev model's ground state by applying a loop gas (LG) projector \citep{spin_one_half,2020-spin-one-kitaev} on the state generated by imaginary time evolution (ITE) \cite{simple-update}. 
We identify the nature of $\mathbb{Z}_2$ QSL of the spin-$1$ Kitaev model by evaluating the virtual order parameters naturally defined in $\mathbb{Z}_2$-invariant PEPS \citep{2020_order_parameter}. 
Due to the fundamental distinction between the integer and half-integer LG projector \citep{lee2020anisotropy}, we found that in contrast to the dispersive Majorana excitation in $S$ = $1/2$, the spin-$1$ Kitaev model possesses a dispersive charge {anyon} excitation.
Also, the TM spectrum suggests the existence of the charge excitations with a small gap at the $\mathbf{\Gamma}$,$\mathbf{K}$ and $\mathbf{K'}$ points in the Brillouin zone.
Minima of  the two-particle continuum are  identified by the excitations {belonging to the vacuum sector} which correspond to the two-particle charge excitation. 
%

\begin{figure}[t]
 \centering
\includegraphics[width=\linewidth]{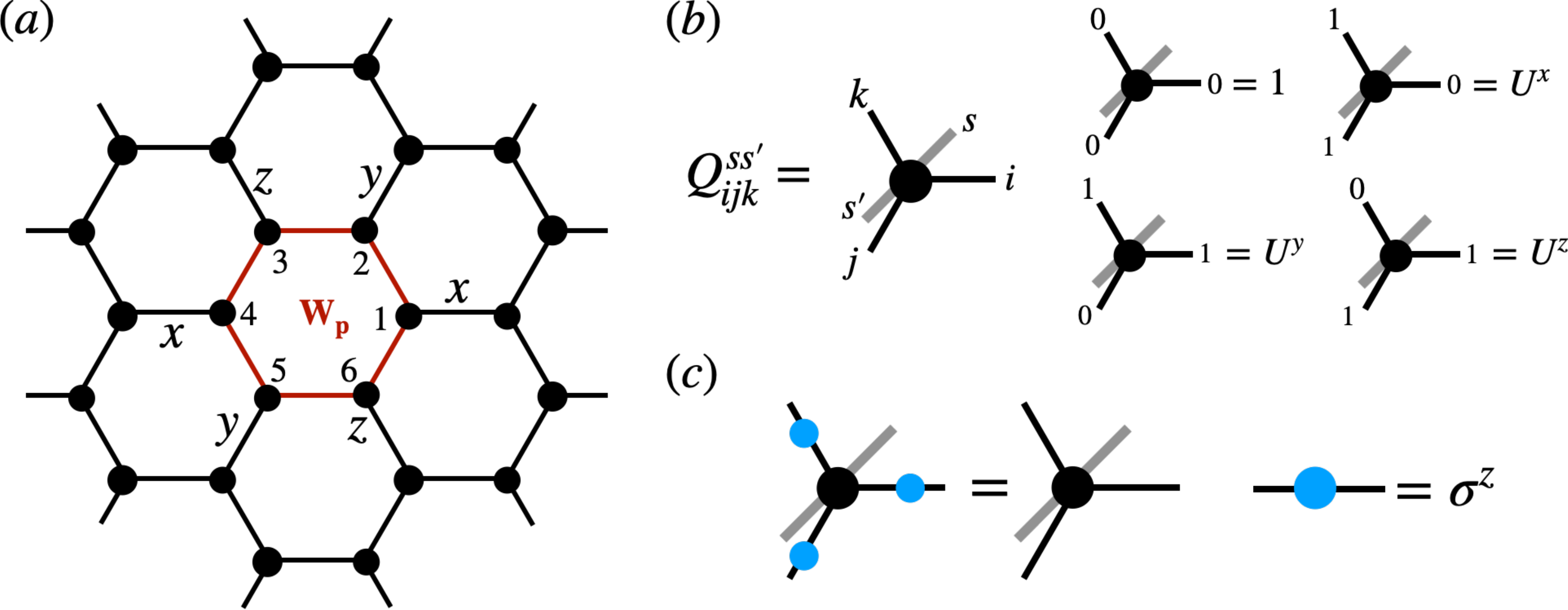}
\caption{(a) Honeycomb lattice and the flux operator $W_p$. (b) Definition of the LG tensor. (c) The LG tensor is invariant under the global $ \sigma^z$ action on the virtual Hilbert space.} 
\label{fig:model_and_LG}
\end{figure}

The honeycomb Kitaev model is given by 
\begin{equation}
H = -J_x \sum_{ \langle i,j  \rangle_x } S_i^xS_j^x - J_y \sum_{\langle i,j \rangle_y} S_i^yS_j^y  - J_z\sum_{\langle i,j \rangle_z} S_i^zS_j^z,
\end{equation}
where $\langle i,j \rangle_\gamma$ represents the nearest neighboring sites connecting through $\gamma$-links where $\gamma = x,y,z$ [Fig.~\ref{fig:model_and_LG}(a)]. 
The flux operator $W_p = U_1^xU_2^yU_3^z U_4^xU_5^yU_6^z$, where $U^\gamma = e^{i\pi S^\gamma}$, is a constant of motion and commutes with the Hamiltonian [Fig.~\ref{fig:model_and_LG}(a)]. 
Here $p$ denotes the site label of the hexagonal plaquettes. 
The Hilbert space is thus divided into sectors according to  the eigenvalues of the flux operator $w_p=\pm 1$.
For the spin-1 Kitaev model, it has been numerically shown that the ground state lives in the vortex-free sector $w_p = +1,\ \forall p$ \citep{spin_one_half, lee2020anisotropy}.
Therefore, the states with $w_p = -1$ for a plaquette $p$ can be understood as a $\mathbb{Z}_2$ vortex quasiparticle.  

From now on, we focus on the isotropic case: $ J_x = J_y = J_z=1$. 
To obtain the tensor network wavefunctions, let us first consider an LG operator
 ${Q}_{\text{LG}} = \text{tTr}\prod_\alpha {Q}^{ss'}_{i_\alpha j_\alpha k_\alpha}|s\rangle \langle s'| $ with the  non-zero elements of the LG tensor defined as
\begin{equation}
Q_{000} = \mathbb{I}, Q_{011} = U^x, Q_{101} = U^y, Q_{110} = U^z.
\end{equation}
where $U^\gamma = e^{i \pi S^\gamma}$ is the $\pi$-rotation operator for a given spin [Fig.~\ref{fig:model_and_LG}(b)]~\cite{lee2020anisotropy}.
${Q}_{\text{LG}} $ is a projector to the vortex-free space: ${W}_p{Q}_{\text{LG}} = {Q}_{\text{LG}}{W}_p = {Q}_{\text{LG}}$. \citep{supp}.
By construction, the LG tensor is invariant under the global $ \mathbb{Z}_2$ symmetry on the virtual Hilbert space: $Q(u_g \otimes u_g \otimes u_g) = Q $ with {$ g \in \{ I,Z \}$ and $u_I =\mathbb{I}_2, u_Z =  \sigma^z $} [Fig.~\ref{fig:model_and_LG}(c)].
This makes a tensor applied by the LG tensor $\tilde{T}^{s}_{i i_0,j j_0,k k_0}= \sum_{s'} Q^{ss'}_{i,j,k}T^{s'}_{i_0,j_0,k_0}$ also has the virtual  $\mathbb{Z}_2$ symmetry with $u_I = \mathbb{I}_2 \otimes_{\alpha} \mathbb{I}_{D_0}, u_Z =  \sigma^z \otimes \mathbb{I}_{D_0} $, where $D_0$ is the bond dimension of the original tensor $T$.
The wave functions formed by contracting the virtual bonds of $\tilde{T}$ is then called a $\mathbb{Z}_2$-invariant PEPS \citep{2011_Norbert_Ginjective,Schuch_2012}.
To simplify the notation, we denote {$u_I(u_Z)$} as ${I}$($Z$) and the operator that anti-commutes with $Z$ as $X$ if there is no ambiguity.
The $\mathbb{Z}_2$-invariant PEPS forms a natural framework to describe $\mathbb{Z}_2$ QSLs. 
For instance, one can construct a flux anyon by attaching half-infinite $Z$ string on virtual bonds and a charge anyon by attaching a single $X$ on virtual bonds based on the notion of the parent Hamiltonian \citep{2017_anyon_condensates}.
%
%

Furthermore, one can show that the flux anyon built on top of $\tilde{T}$, i.e., any tensor applied by the LG tensor, corresponds to a single $\mathbb{Z}_2$ vortex $w_p = -1$ at the starting point of the half-infinite $Z$ string  ~\citep{2020-spin-one-kitaev}. 
Similarly, the single $X$ creating charge anyon on site $i$ correspond to a physical half-infinite string operator $ \prod_{n = i}^{\infty} U^{\gamma_{n}}_n$. Here $(i,i+1,...)$ are the sites for the string and $(\gamma_i,\gamma_{i+1},...)$ are the links normal to the string (See the Supplemental Material~\citep{supp} for details). 
It is interesting to note that this operator is the same as the disorder operator defined in Ref.~\citep{Baskaran_2008} up to a phase factor.
Note that a $\mathbb{Z}_2$-invariant PEPS does not necessarily guarantee a $\mathbb{Z}_2$ topologically ordered phase, for the system can be driven into a trivial \citep{Norbert_Schuch_2013,Haegeman_2015} {or even a non-Abelian \citep{non-AbelianTO_2020, our_paper}} phase by a physical deformation of the local tensor. 
This property makes it as a suitable ans\"atz to study whether the system harbors a QSL phase and identify possible topological transitions \citep{non-AbelianTO_2020,He_2014,Mei_2017}.

%

To obtain a $\mathbb{Z}_2$ invariant ground state, we first perform the ITE  and then apply the LG projector on the resulting states, i.e., $|\psi \rangle = \lim_{\tau \rightarrow \infty }{{Q}_{\text{LG}}}e^{-\tau H}|\psi_0 \rangle$. 
{
Here, the initial product state $|\psi_0 \rangle = \otimes_{\alpha} |(111)\rangle_\alpha$, where $|(111)\rangle$ is the magnetized state along $(1,1,1)$ direction: $\langle (111)|\Vec{S}|(111)\rangle = (1,1,1)/\sqrt{3}$.}
This ensures that the PEPS with a fixed bond dimension $D$ is both $\mathbb{Z}_2$-invariant and vortex-free. 
This also allows for a lower computational cost than that of  the gauge-symmetry-preserved update  \citep{He_2014,Mei_2017}, as only half of the bond dimensions is needed during the ITE process.
Since the two-site time-evolution operator $e^{-\tau S_i^\gamma S_j^\gamma}$  commutes with  ${Q}_{\text{LG}}$,
one can numerically show that ${Q}_{\text{LG}}e^{-\tau H}|\psi_0 \rangle$ acquires equal energy as $e^{-\tau H}|\psi_\text{LG} \rangle = e^{-\tau H}{Q}_{\text{LG}}|\psi_0 \rangle$ for the same bond dimension.

\label{sec:TM}

\begin{figure}[t]
 \centering
\includegraphics[width=\linewidth]{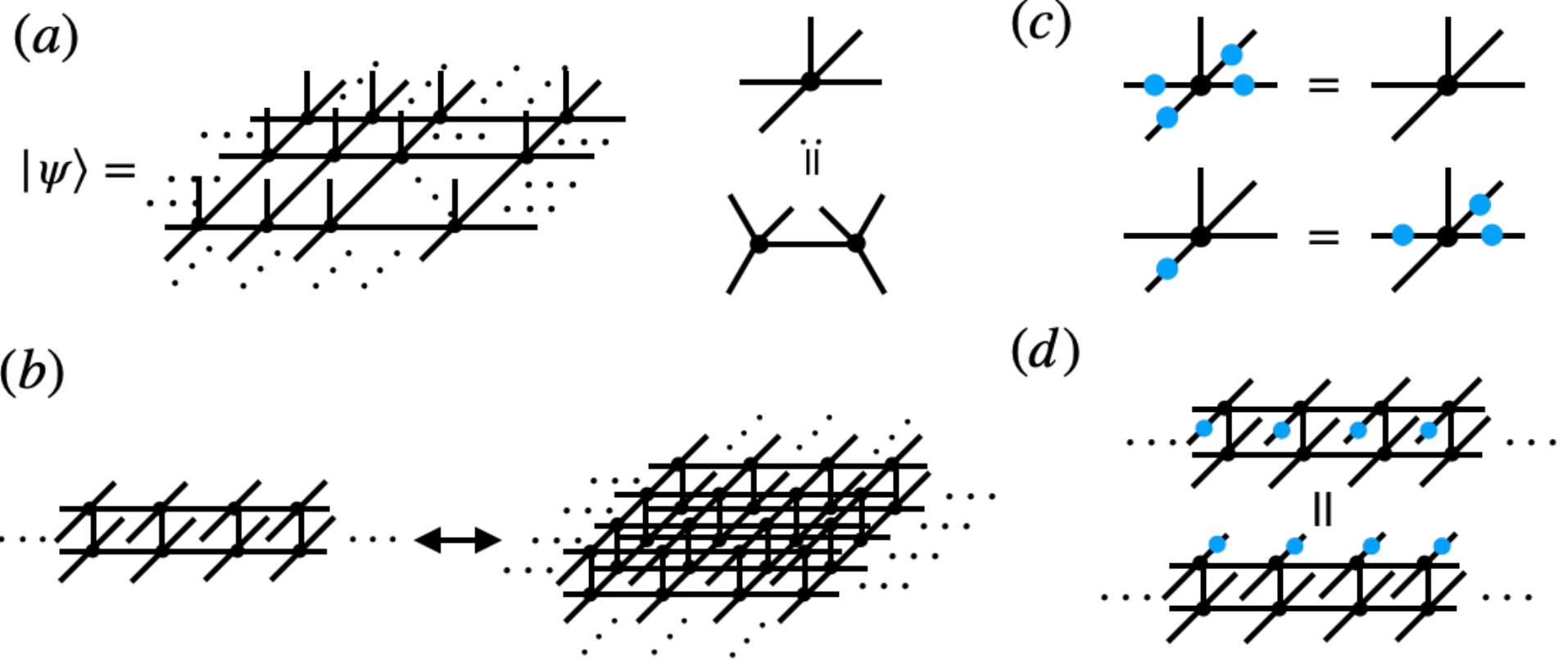}
\caption{(a)  2D tensor network wave function. Tensors on the two sublattices of the honeycomb lattice are merged into a single tensor. (b) Norm of an infinite 2D PEPS wave function (right) can be constructed by repeating the 1D TM (left). (c) $\mathbb{Z}_2$-invariant property of the tensor and the resulting pulling through condition. Here we use blue circle to denote the group action $u_g$ with $g \in \{I, Z\}$. (d) Schematic representation of TM's symmetry inherited from the $\mathbb{Z}_2$-invariant tensor: $[\mathbb{T}, Z^{\otimes \infty} \otimes \mathbb{I}] = 0$. Another TM's symmetry $[\mathbb{T},  Z^{\otimes \infty}\otimes Z^{\otimes \infty} ] = 0$ can be derived in a similar way.} 
\label{fig:TM_and_sym}
\end{figure}

Once the PEPS wave function is obtained [Fig.~\ref{fig:TM_and_sym}(a)], the topological property and low lying excitations can be studied by considering its transfer matrix (TM) $\mathbb{T}$ [Fig.~\ref{fig:TM_and_sym}(b)] \citep{2015_Zauner}. 
The TM can be regarded as the building block of the norm of the two-dimensional PEPS: $\langle \psi |\psi \rangle = (l| \lim_{L_y \rightarrow \infty} \mathbb{T}^{L_y} |r) = (l| \big[ \lim_{L_y \rightarrow \infty} |l)\lambda^{L_y}(r| \big] |r) $. 
Here $|l)$($|r)$) is the left(right) dominant eigenvector of $\mathbb{T}$ and $\lambda$ is the corresponding dominant eigenvalue which is normalized to $1$ (Here we use $|\cdot )$ to denote the vector defined in the virtual Hilbert space).
In our case, $|l)$ and $|r)$ are the same and we denote them as $|\rho)$ in the following.

\begin{table}
\caption{
Relation between three different entanglement phases, virtual order parameters, and  symmetries of the fixed point.
}
\begin{tabular}{ ||c|c|c|c || } 
 \hline 
            & \text{ Confinement } & \text{ Topological Order } & \text{ Condensation } \\ 
 \hline 
  \makecell{$(\rho|X \otimes X|\rho)$\\ $ Z^{\otimes \infty}\otimes \mathbb{I}^{\otimes \infty}$  } & \makecell{0 \\ preserved} & \makecell{$\neq 0$ \\ broken} & \makecell{$\neq 0$ \\ broken} \\ 
  \hline
 \makecell{$(\rho|X \otimes I|\rho)$\\ $ Z^{\otimes \infty}\otimes Z^{\otimes \infty}$  } & \makecell{0 \\ preserved} & \makecell{0 \\ preserved} & \makecell{$\neq 0$ \\ broken}\\ 
 \hline
\end{tabular}
\label{table:VOPs}
\end{table}

Due to the $\mathbb{Z}_2$-invariance, $\mathbb{T}$ inherits the symmetry from the tensor: $ [\mathbb{T}, Z^{\otimes \infty} \otimes \mathbb{I}^{\otimes \infty}] = [\mathbb{T},  Z^{\otimes \infty}\otimes Z^{\otimes \infty} ]  = 0$ [Fig.~\ref{fig:TM_and_sym}(d)] through the pulling through condition [Fig.~\ref{fig:TM_and_sym}(c)]. 
The entanglement spectrum can be extracted from the fixed point $|\rho)$, which is equivalent to the boundary of a PEPS \citep{2011_entanglement}.  
Interestingly, once the TM has the above symmetries, the fixed point may not be unique, and there are three different \textit{entanglement phases} \citep{2017_anyon_condensates}: 
(1) The fixed point is unique and respects both the $ Z^{\otimes \infty}\otimes Z^{\otimes \infty} $ and $Z^{\otimes \infty} \otimes \mathbb{I}^{\otimes \infty}$ symmetries. 
(2) The fixed point is two-fold degenerate and respects the $Z^{\otimes \infty}\otimes Z^{\otimes \infty} $ symmetry, but breaks the $Z^{\otimes \infty} \otimes \mathbb{I}^{\otimes \infty}$ symmetry. 
(3) The fixed point is four-fold degenerate and breaks both the $ Z^{\otimes \infty}\otimes Z^{\otimes \infty} $ and $Z^{\otimes \infty} \otimes \mathbb{I}^{\otimes \infty}$ symmetries. 
To detect whether the fixed point respects the symmetry, one can borrow the notion of Landau's symmetry breaking paradigm by measuring the expectation value of the virtual order parameters $ X \otimes X $ and $ X \otimes \mathbb{I} $ which anti-commutes with $ Z^{\otimes \infty}\otimes \mathbb{I}^{\otimes \infty} $ and $Z^{\otimes \infty} \otimes Z^{\otimes \infty}$, respectively (Tab.~\ref{table:VOPs}). 
The connection between the entanglement and the physical phases lies in the fact that the single $X$ action on a virtual bond of a $\mathbb{Z}_2$-invariant PEPS corresponds to the creation of a charge anyon. 
As a consequence, the virtual order parameters $(\rho| X \otimes X|\rho)$ and $(\rho| X \otimes \mathbb{I}|\rho)$ are equivalent to the overlap of the physical wave functions $\langle \psi_{e} | \psi_{e} \rangle$ and $\langle \psi_{e} | \psi \rangle$, respectively, where $| \psi_{e}\rangle$ is the charge excited states.
If $\langle \psi_{e} | \psi \rangle  = 1$, the operator creating charge actually does nothing to the ground state, and thus the charge anyon is condensed. On the other hand, if the charge excited state is not  a properly normalizable quantum state, i.e., $ \langle \psi_{e} | \psi_{e} \rangle  = 0$, the system is in a charge confined phase \citep{Haegeman_2015, 2017_anyon_condensates, 2020_order_parameter}.
Therefore, only the second case corresponds to the topologically ordered phase (Tab.~\ref{table:VOPs}). 
In fact, since away from the renormalization group fixed point, a locally created anyon  will propagate, and nonzero overlap $\langle \psi_e |\psi \rangle$ is enough to guarantee a charge condensed phase~\citep{2020_order_parameter}. 
To evaluate the order parameter in the infinite two-dimensional tensor network, we employ the variational uniform matrix product state (VUMPS) algorithm \citep{2018_vumps, 2018_faster_method, 2019_tangent_space_lecture} whose accuracy can be controlled by the bond dimension of MPS $D_\text{mps}$ (See the Supplemental Material~\citep{supp} for details ).
Throughout the calculation, we find that $(\rho| X \otimes X |\rho) = 1$ and $(\rho| X \otimes \mathbb{I} |\rho) = 0$ regardless of the bond dimension (up to $D = 10$), suggesting that the ground state of spin-1 Kitaev model lies in the $\mathbb{Z}_2$ spin liquid phase. 
This method has an advantage over  identifying QSL phase by  the topological entanglement entropy, which is limited by the small bond dimension and suffers from the finite-size effect \citep{2020-spin-one-kitaev}.

The TM's sub-dominant eigenvalues  encompass signatures of  the low-energy excitations~\citep{2015_Zauner, Haegeman_2015, He_2017, 2018_Kitaev_QSL,Hu_2019}. 
This is a manifestation of the fact that the information of a local Hamiltonian's excitations is encoded in the ground state, which can be extracted by measuring the ground state correlations.
The prominent example is that the minus logarithm of the largest sub-leading eigenvalue $ -\log{|\lambda|} $, which corresponds to  the inverse of the correlation length, can be related to the spectral gap up to an overall energy scale \citep{Hastings_2004}. 
This argument has been further extended in Ref.~\cite{2015_Zauner} to include the momentum dependence. 
To be more specific, for a generally complex eigenvalue $\lambda = e^{-\epsilon +i\phi}$ of the TM, the corresponding physical excitation energy is given as $E \sim \epsilon = -\log{|\lambda|}$, while the corresponding momentum is related to the phase $k \sim \phi = \arg \lambda$.
Therefore, by solving the eigenvalue problem of the \textit{transfer matrix Hamiltonian} $H_{\mathbb{T}} = -\log{\mathbb{T}}$, one can access the physical Hamiltonian's low-lying excitations. 
The fixed point $|\rho )$ now correspond to the ground state of the TM Hamiltonian $H_{\mathbb{T}}$ since the energy $\epsilon = -\log|\lambda| = 0$.
Since the TM Hamiltonian is one-dimensional, the excitations for this Hamiltonian, which we term TM excitations, can be studied by constructing the excitation ansatz  on top of the fixed point \citep{supp,2012_excitation, 2019_tangent_space_lecture}.
%
%
%
%

%
%

Symmetry of the fixed points implies the distinct types of TM excitations.
Since the topological phase preserves $Z^{\otimes \infty} \otimes Z^{\otimes \infty}$ symmetry (Tab.~\ref{table:VOPs}), the TM excitations can be characterized by the eigenvalues of $ Z^{\otimes \infty} \otimes Z^{\otimes \infty} $,
which can either be $+1$ or $-1$. 
Due to  the symmetry breaking of $Z^{\otimes \infty} \otimes \mathbb{I}^{\otimes \infty}$ (Tab.~\ref{table:VOPs}), the domain-wall excitation of the TM, i.e., the interpolation between two degenerate fixed points, should also be considered. 
Overall, it follows that there are four different TM excitations. 
The connection between the physical excitations and different types of TM excitations can be understood as follows.
Since the operator $Z^{\otimes \infty}$ acting on virtual bonds anti-commutes with the action $X$ creating charge anyon, it can be regarded as the Wilson loop operator detecting physical charge. 
Therefore, $Z^{\otimes \infty} \otimes Z^{\otimes \infty} $ corresponds to a physical operator measuring charge difference between the ket and bra layers. 
Similarly, as a single flux anyon can be created by attaching a half infinite $Z$ string, the interpolation between different fixed points breaking $Z^{\otimes \infty} \otimes \mathbb{I}^{\otimes \infty} $ corresponds to creating flux on either the ket or bra layer. 
As a consequence, the trivial (i.e., no domain wall) TM excitations with $Z^{\otimes \infty} \otimes {Z}^{\otimes \infty} = +1$ correspond to the  vacuum-sector excitations, while the trivial TM excitation with $Z^{\otimes \infty} \otimes {Z}^{\otimes \infty} = -1$ corresponds to the charge-sector excitations.
Similarly, the domain-wall TM excitations with $Z^{\otimes \infty} \otimes {Z}^{\otimes \infty} = +1$ and $-1$  are related to the physical flux- and fermion- sector excitations, respectively. 
In fact, the above argument can be made even more rigorous by using the pulling through condition in Fig.~\ref{fig:TM_and_sym}(c) to show ~\cite{Haegeman_2015} that those four different TM excitations defined through the infinite plane correspond to the regular and mixed TMs \citep{2015_Zauner} defined within the setting of long cylinder.
Interestingly, we find that the LG projector for the integer spins only supports dispersive excitations belonging to the vacuum- and charge- sectors, while the flux- and fermion- sector excitations are static.
This is intrinsically different from the half-integer LG projector, where only vacuum and fermion anyon excitations are dispersive~\citep{our_paper}. 
Interestingly,  this is consistent with the argument put forth in Ref.~\citep{Baskaran_2008} that the integer-spin Kitaev model has bosonic exciations instead of Majorana fermions, indicating that 
the different sign structures of the integer and half-integer spin LG projectors may faithfully describe the distinct nature of the integer and half-integer spin Kitaev models \citep{lee2020anisotropy}.

\begin{figure}[tb]
 \centering
\includegraphics[width=\linewidth]{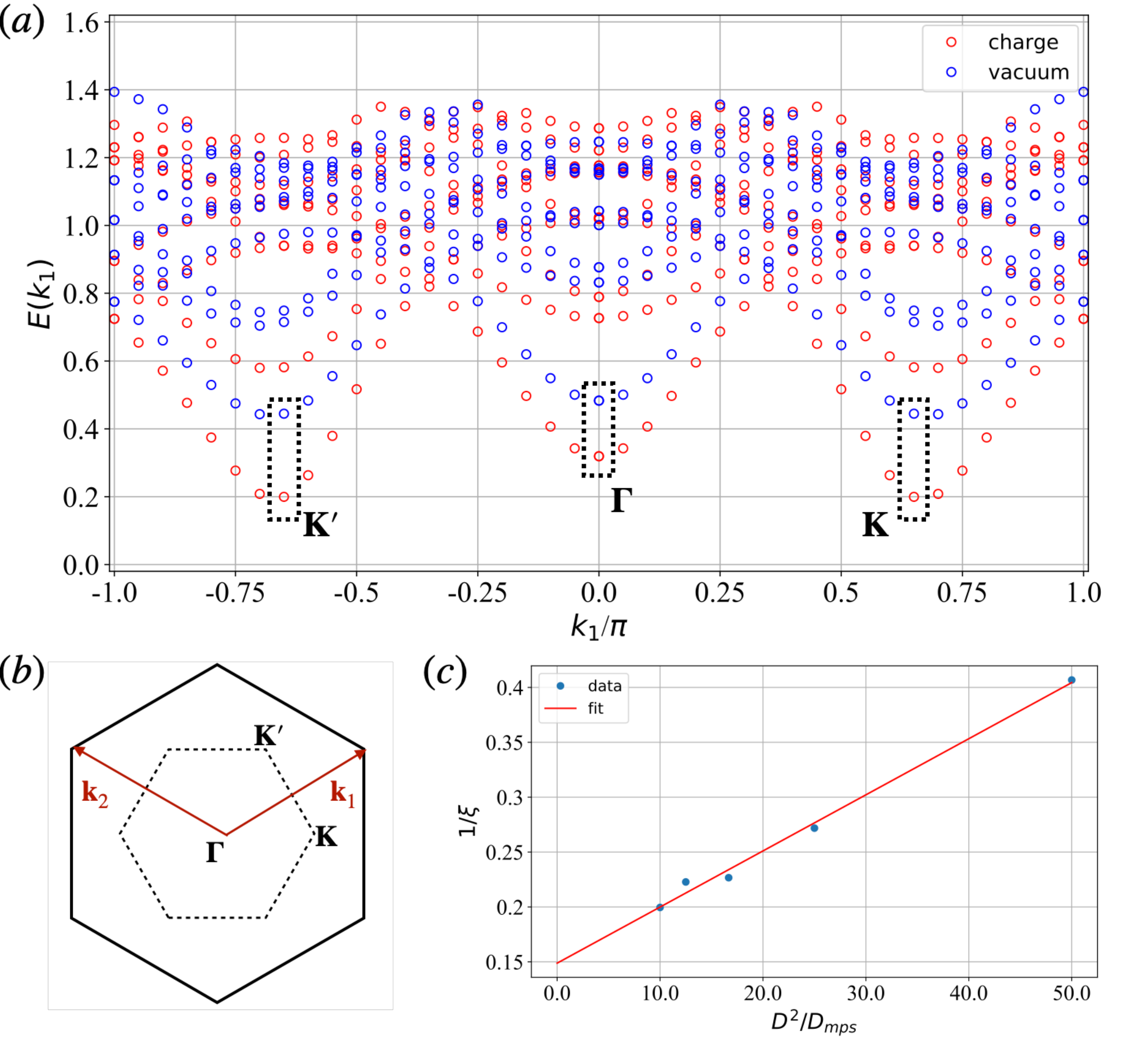}
\caption{(a) Transfer matrix spectrum for the PEPS tensor with $(D,D_\text{mps}) = (10, 10)$. (b) Brillouin zone with labeled positions of the minimum of TM spectrum. (c) The inverse of the correlation length of the $D = 10$ PEPS wavefunction as a function of $D_\text{mps}$.} 
\label{fig:D10spect_BZ_extrapolate}
\end{figure}

Using the correspondence $E_{k_1} \sim -\log|\lambda_{k_1}|$ for a given momentum $k_1$ in the $x$-direction,
the TM spectrum for PEPS states with bond dimension $D=10$ are shown in Fig.~\ref{fig:D10spect_BZ_extrapolate}(a).
The excitations belonging to charge (red) and vacuum (blue) sectors are distinguished by calculating the charge difference $(\rho_\text{ex}|Z^{\otimes \infty}\otimes Z^{\otimes \infty}|\rho_\text{ex})$, where $|\rho_{ex})$ is the TM excitations.
While overall energy scale of the local minimum of $E_{k_1}$ and exact excitation energies are unknown due to the lack of the knowledge of the Lieb-Robinson velocity \citep{1972_Lieb}, the corresponding momentum $k_1$ at the local minimum of $E_{k_1}$ allows us to identify the location of the low-energy dispersion \cite{Hastings_2004}. 
It then follows that both the charge- and vacuum- sector excitations are clearly identified at $k_1 = 0, 2\pi/3,$ and $-2\pi/3$.
We also perform the corner transfer matrix renormalization group (CTMRG) \citep{doi:10.1143/JPSJ.65.891,PhysRevB.80.094403,PhysRevB.82.245119} to obtain the TM spectrum (See the Supplemental Material~\citep{supp} for details ), and the results are consistent. 
To gain more insights into the two-dimensional system, we consider the momentum in the $y$-direction using $k_2 \sim \arg \lambda_{k_1}$.
%
The location of three minimum excitations are then identified at $(k_1,k_2) = (0,0), (2\pi/3, -2\pi/3),$ and $(-2\pi/3, 2\pi/3)$, suggesting that the spin-$1$ Kitaev model harbors three low-lying charge anyon excitations at  $\mathbf{\Gamma}, \mathbf{K},$ and $\mathbf{K'}$  points in the Brillouin zone [Fig.~\ref{fig:D10spect_BZ_extrapolate}(b)].
To nail down the origin of the vacuum-sector excitations, which is possible to be all the even-particle excitations, we note that the global minima lie at $\mathbf{K} (\mathbf{K'})$: $\epsilon_\text{min} = \epsilon(\mathbf{K}) = \epsilon(\mathbf{K'}) <  \epsilon(\mathbf{\Gamma})$.
The low-lying vacuum-sector excitations at  $\mathbf{\Gamma}, \mathbf{K},$ and $\mathbf{K'}$ can then be well explained by attributing to the two-particle charge excitations $\epsilon(\mathbf{K}) + \epsilon(\mathbf{K'})$, $\epsilon(\mathbf{K'}) + \epsilon(\mathbf{K'})$, and $\epsilon(\mathbf{K}) + \epsilon(\mathbf{K})$, respectively. 
Therefore, we conclude that the excitations belonging to the vacuum sector describe the minima of the two-particle continuum.

Figure~\ref{fig:D10spect_BZ_extrapolate}(c) shows the the inverse of the correlation length, i.e., the corresponding charge excitation energy at $\mathbf{K}(\mathbf{K'})$, as a function of the accuracy-controlled dimension $D_\text{mps}$. 
Extrapolation to  $D_\text{mps}\to \infty $ shows that the system is gapped with a  correlation length of $\xi \approx 6.7$ unit cells.
Using the excitation gap   $E_\text{24site} \approx 3.6 \times 10^{-2}$ from  a $24$-site  exact diagonalization \citep{2018_exact_spin1} as the upper bound, we estimate the characteristic velocity $v_\text{LR} \approx 2.4\times 10^{-1}$ using the relation $\xi E_\text{24site} = v_{\text{LR}}$ \citep{Hastings_2004,2015_Zauner}.
However,  the possibility of a gapless spin liquid cannot be completely ruled out from the current numerics.
%

In this paper, we construct a $\mathbb{Z}_2$-invariant PEPS to study the excitation of the isotropic spin-$1$ Kitaev model. 
Different from the $S$ = 1/2 case where  dispersive Majorana fermion excitations exist, the $S$=1 Kitaev model has  dispersive charge excitations.
We locate the single- and two-particle excitations' minima at $\mathbf{\Gamma}$,$\mathbf{K}$, and $\mathbf{K'}$  for the spin-$1$ Kitaev model. 
%
%
Note that dynamical spin structure factor, which allows direct comparison with the INS experiment, involves  not only the two-particle excitation, but also the static gauge flux as the spin-flip operator will necessary induce a flux anyon pair \citep{Baskaran_2007, Knolle_2014,Knolle_2015}. 
This makes the connection between the INS experiments with the two-particle excitations less trivial.
However, we note that the resonant inelastic X-ray scattering experiments may provide a route to  single out the Majorana sector without the influence of flux in the spin-1/2 Kitaev model~\citep{Hal_sz_2016}. 
It is interesting to further investigate whether a similar scheme can be applied to the spin-$1$ case to detect the signal from the charge sector only.
On the other hand, the construction of the $\mathbb{Z}_2$-invariant PEPS using the LG projector is an efficient method to separate different anyonic excitations and can be easily generalized to the anisotropic Kitaev model. 
By tracing the evolution of the excitation spectra, one should be able to understand whether the QSL feature in the isotropic Kitaev model  persists when the system is driven away from  the isotropic point.
Further studies along these directions are worth pursuing.

\acknowledgements

This work is partially supported by the Ministry of Science and Technology (MOST) of Taiwan under grants No. 107-2112-M-002-016-MY3, and 108-2112-M-002-020-MY3 (YHC, JZ, YJK).
YBK is supported by the NSERC of Canada and the Center for Quantum Materials at the University of Toronto.

\appendix

\section{$\mathbb{Z}_2$-invariant PEPS}

In this section, we give a brief introduction to $\mathbb{Z}_2$-invariant PEPS and how it is applied  to the spin-$1$ Kitaev model. 
We refer interested readers to Refs.~\cite{2011_Norbert_Ginjective,2017_anyon_condensates} for details.

\subsection{$\mathbb{Z}_2$-invariant PEPS, degenerate ground states, and anyons}
\begin{figure}[b]
\centering
\includegraphics[width=\linewidth]{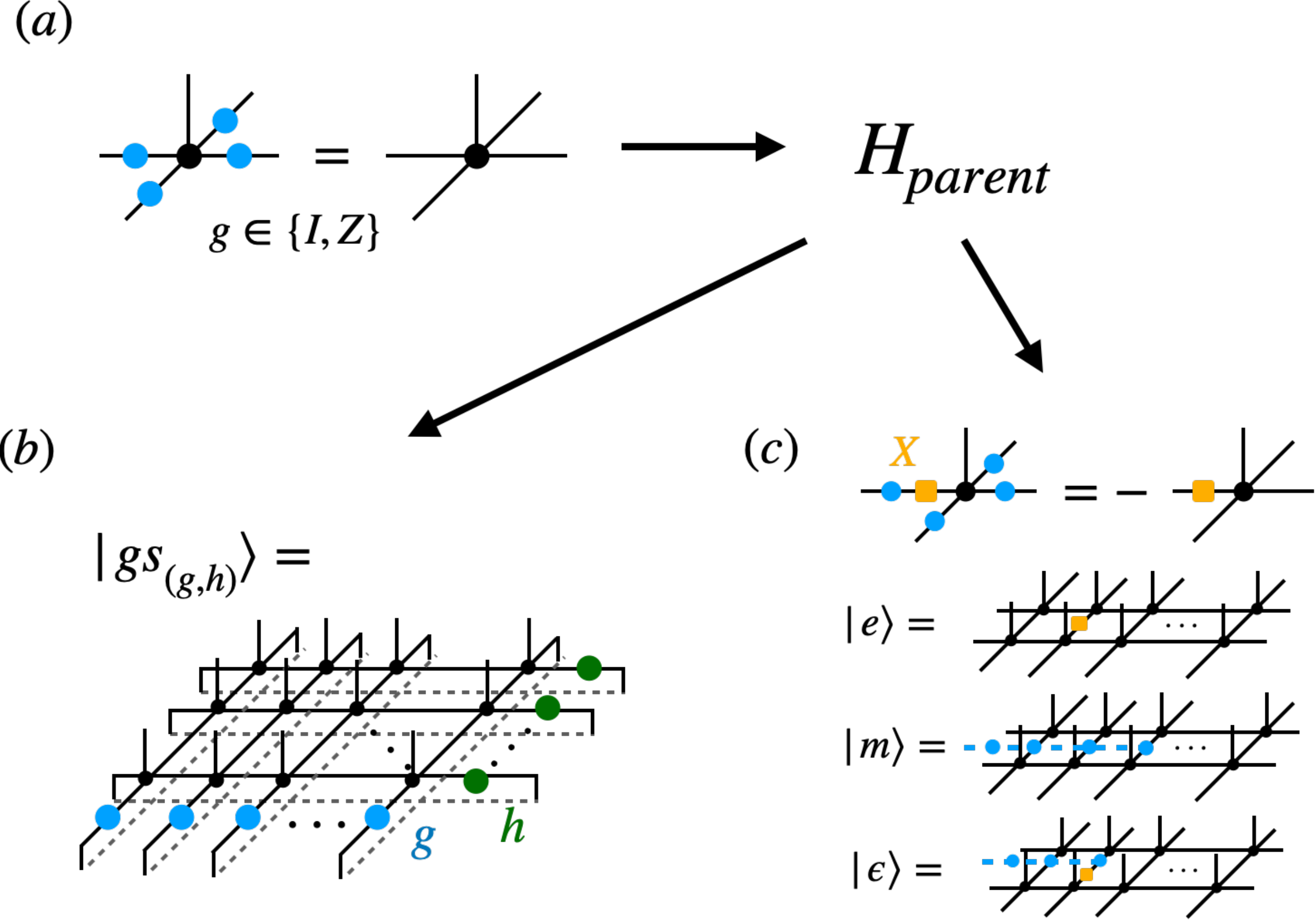}
\caption{(a) Given a $\mathbb{Z}_2$-invariant PEPS, one can construct a parent Hamiltonian which is a sum of local terms. (b) The ground states subspace of the parent Hamiltonian is spanned by $|{gs}_{(g,h} \rangle$ with $g,h \in \{I,Z\}$. (c) The charge anyon can be crated by applying a single $X$ action on the virtual bond. The flux anyon can be created by a half-infinite $Z$ string. The fermion can be constructed by applying both the $X$ and half-infinite $Z$ string. } 
\label{fig:Z2-PEPS}
\end{figure}


$\mathbb{Z}_2$-invariant PEPS forms a natural framework to describe $\mathbb{Z}_2$ QSLs.
As shown in Ref.~\citep{2011_Norbert_Ginjective}, given a $\mathbb{Z}_2$-invariant PEPS, one can construct a parent Hamiltonian which is a sum of local terms [Fig.~\ref{fig:Z2-PEPS}(a)], such that its ground state subspace  is spanned by two non-contractible loop operators on the torus [Fig.~\ref{fig:Z2-PEPS}(b)]. 
The reason for the degenerate energy of the four states $|{gs}_{(g,h)}\rangle$ with $g,h \in \{I,Z\}$ is that the parent Hamiltonian cannot distinguish those states locally, as the loop operators can be deformed to other places using the pulling through condition discussed in Fig.2(c) of the main text. 
Furthermore, we can create anyonic excitations of the parent Hamiltonian on top of any four ground states [Fig.\ref{fig:Z2-PEPS}(c)] \citep{2017_anyon_condensates}. The charge anyon $|e\rangle$ can be created by attaching a single $X$ on a virtual bond of the $\mathbb{Z}_2$-invariant PEPS. 
Since the resulting tensor transforms anti-invariant instead of invariant under the global $\mathbb{Z}_2$ action, one can show that no operator acting on the physical Hilbert space can create a single charge anyon, which is a feature of topologically non-trivial excitations.
On the other hand, the flux anyon $|m\rangle$ can be created by a half-infinite $Z$ string, as the starting point of the string cannot be moved and hence will be detected by the parent Hamiltonian. 
Interestingly, the $\mathbb{Z}_2$ anti-invariant condition of the charge anyon now corresponds to the braiding rule between the charge and flux anyons.
Finally, the fermion $|\epsilon\rangle$ can be created using the braiding rule $\epsilon = e\times m$ of the $\mathbb{Z}_2$ QSLs.

Note that away from the renormalization group fixed point, the excitations are dispersive, and the anyonic states discussed above are not the eigenstates of the parent Hamiltonian. Instead, the eigenstates should be solved by the superposition of these states. However, the construction in Fig.~\ref{fig:Z2-PEPS}(c) gives the correct quasiparticle information and can still be used to extract low-energy dispersion from the TM \citep{Haegeman_2015}. 
Also, the formal proof of the existence of parent Hamiltonian requires the tensor to be $\mathbb{Z}_2$-injective, which means that the global $\mathbb{Z}_2$ action on the virtual bonds is the only symmetry of the PEPS tensor.

\subsection{Excitations in Spin-$1$ Kitaev model}
\begin{figure}[b]
\centering
\includegraphics[width=\linewidth]{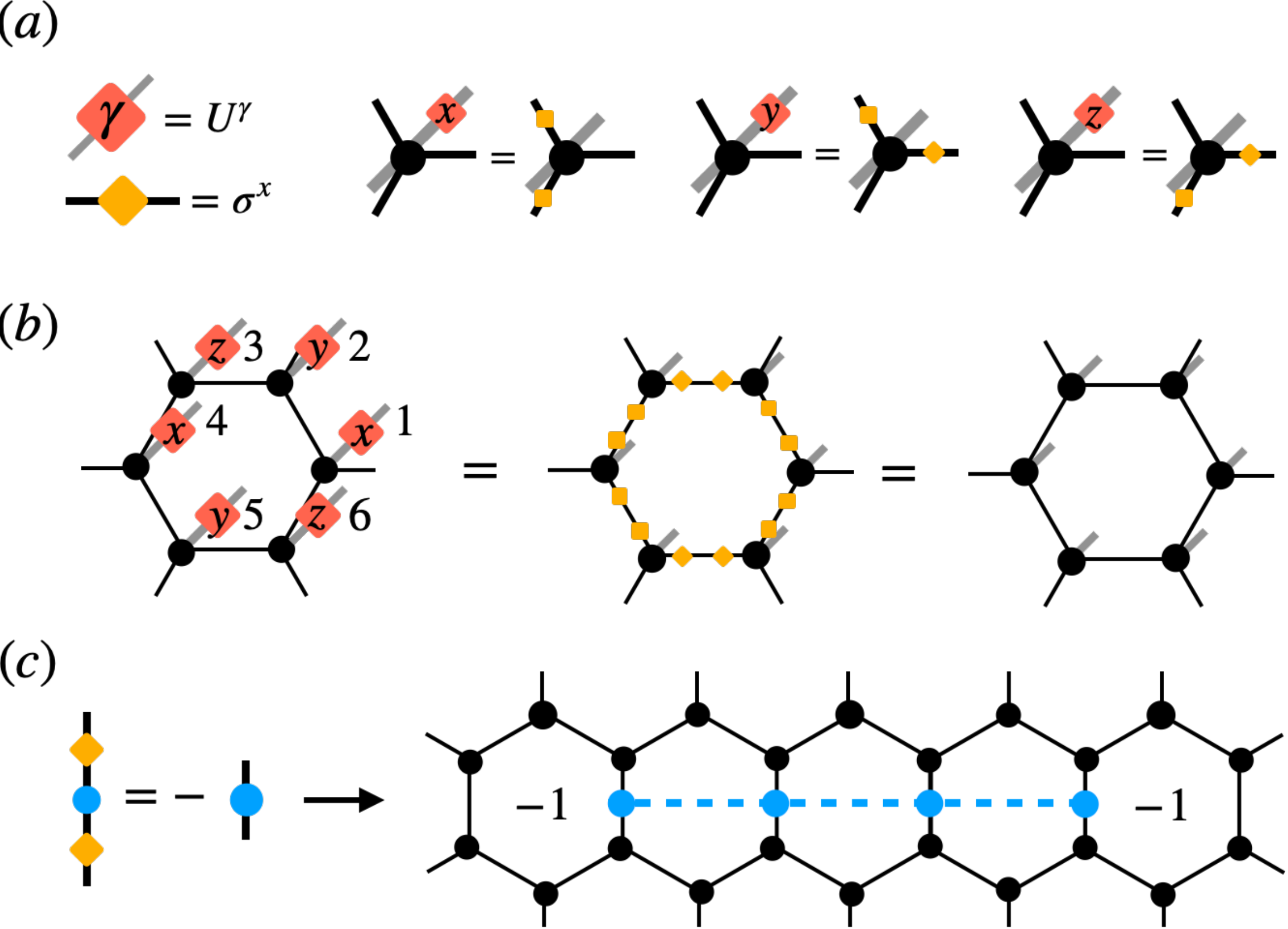}
\caption{ 
(a) The schematic representation of Eq.~\eqref{eqn:supp2}. (b) The action of the flux operator $W_p$ to the LG operator (c) The flux anyon pair in the framework of $\mathbb{Z}_2$-invariant PEPS corresponds to the $\mathbb{Z}_2$ vortex pair of spin-$1$ Kitaev model. Here the physical bonds are suppressed and the lattice is rotated $90$ degrees clockwise.
} 
\label{fig:LG_op}
\end{figure}

To understand the connection between the spin-$1$ Kitaev model and anyons created in $\mathbb{Z}_2$-invariant PEPS framework, we note that the LG tensor satisfies the following relations \citep{2020-spin-one-kitaev}:
\begin{align}
\label{eqn:supp2}
U^x Q_{ijk} &= \sum_{j'k'} \sigma^x_{jj'} \sigma^x_{kk'} Q_{ij'k'},\nonumber\\
U^y Q_{ijk} &= \sum_{k'i'} \sigma^x_{kk'} \sigma^x_{jj'} Q_{i'jk'},\nonumber\\
U^z Q_{ijk} &= \sum_{i'j'} \sigma^x_{ii'} \sigma^x_{jj'} Q_{i'j'k},
\end{align}

\noindent
which can be schematically represented in Fig.~\ref{fig:LG_op}(a). This immediately shows that a single $U^\gamma$ creates a pair of charge excitations, and thus a half-infinite string operator $ \prod_{n = i}^{\infty} U^{\gamma_{n}}_n$ discussed in the main text corresponds to the creation of a single charge anyon.

Using the relation in Fig.~\ref{fig:LG_op}(a), it is straightforward show that ${Q}_\text{LG}$ is a projector to the vortex-free space ${W}_p{Q}_{\text{LG}} = {Q}_{\text{LG}}{W}_p = {Q}_{\text{LG}},\ \forall p$. The schematic proof of the case that $W_p = U_1^xU_2^yU_3^z U_4^xU_5^yU_6^z$ is shown in Fig.~\ref{fig:LG_op}(b).
Similarly, using $\sigma^x\sigma^z\sigma^x = -\sigma^z$, one can show that the flux anyon pair constructed on the virtual bonds correspond to two vortices with $w_p = -1$ [Fig.~\ref{fig:LG_op}(c)].

\begin{figure}[tp]
 \centering
\includegraphics[width=0.9\linewidth]{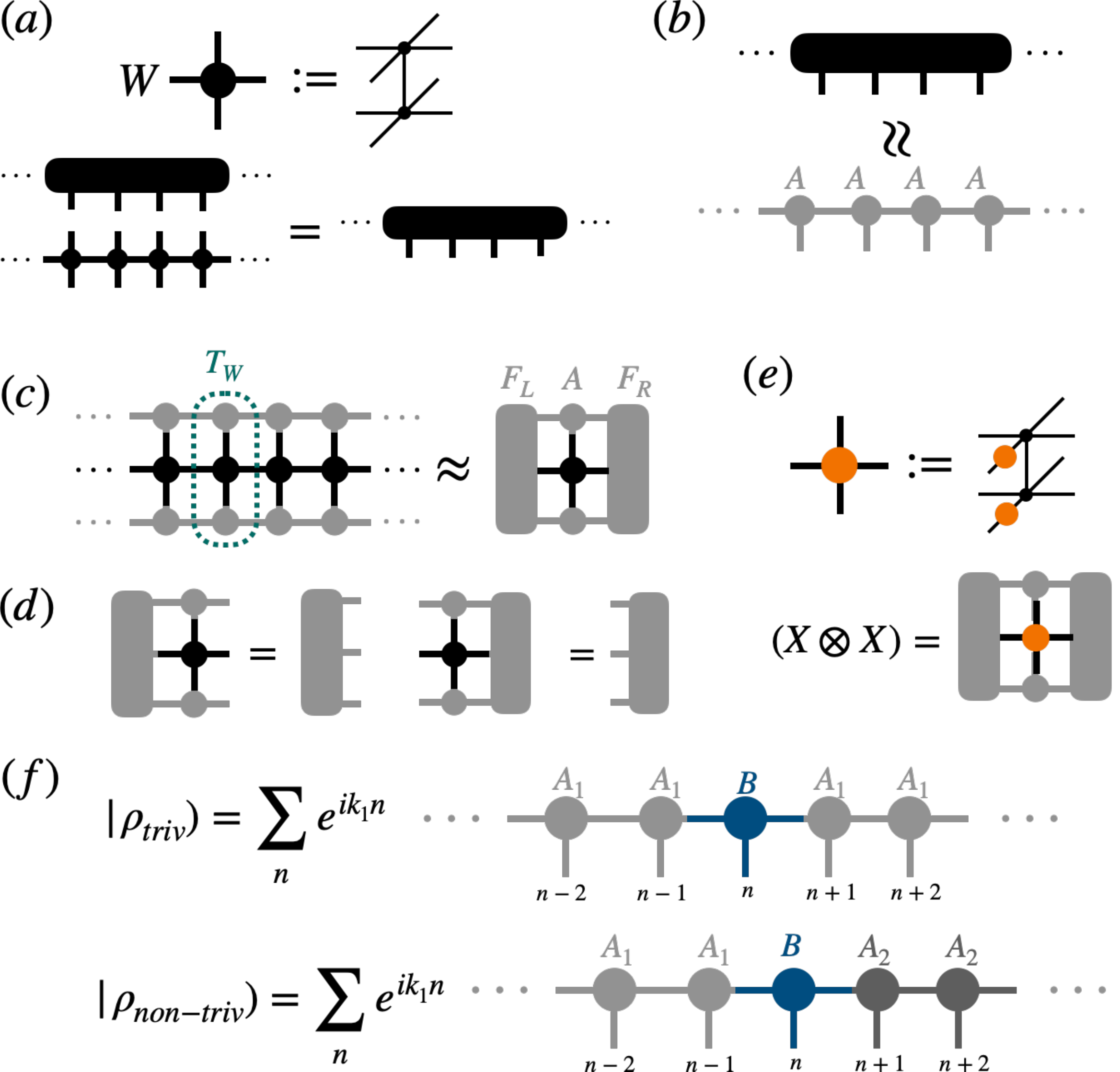}
\caption{
See text for details.
} 
\label{fig:tangent_space}
\end{figure}

\section{VUMPS and Excitation}
\label{sec:tangent-space}
In this section, we give an overview about how to compute environment of the infinite two-dimensional tensor network using VUMPS and then solve the transfer matrix spectrum by constructing the excitation an\"{a}stz based on the fixed point obtained in VUMPS. We refer the interested readers to Refs.~\citep{2018_faster_method,2018_vumps,2019_tangent_space_lecture,2012_excitation} for details.

When it comes to evaluating the observable in the infinite two-dimensional tensor network, the fixed point of the transfer matrix is needed. 
The transfer matrix can be represented by the infinite contraction of the double tensor $W$, i.e., the tensor acquired by tracing out the physical indices [Fig.~\ref{fig:tangent_space}(a)].
Since the system considered is transitionally invariant, we can approximate the fixed point $|\rho)$ as a uniform MPS [Fig.~\ref{fig:tangent_space}(b)]. 
The infinite-dimensional dominant eigenvalue problem is then transformed into finding the optimal MPS such that it maximizes the infinite contraction of the channel operator $T_W$ [Fig.~\ref{fig:tangent_space}(c)], 
which requires the evaluation of the left(right) environment ($F_L(F_R)$) by solving the  left(right) dominant eigenvector the channel operator [Fig.~\ref{fig:tangent_space}(d)]. 
By variationally optimizing the MPS and then updating the  left and right environments, the TM fixed points can be obtained. 
The resulting tensors $A,F_L,$ and $F_R$ can then serve as an environment for any local observable. 
For instance,  we want to evaluate the virtual order parameters $(\rho |X \otimes X| \rho )$ and $(\rho |X \otimes \mathbb{I}| \rho )$ to identify the $\mathbb{Z}_2$ QSL nature of the spin-1 Kitaev model.
The quantity $(\rho |X \otimes X| \rho )$ can be computed by attaching two $X$s on both the bra and ket layers of $W$'s virtual bonds and then contract with the environment [Fig.\ref{fig:tangent_space}(e)]. 
Similar construction can be done for $(\rho |X \otimes \mathbb{I}| \rho )$. 
Physical observables can also be calculated by sandwiching the operator inside the physical bonds of $W$ and then contracting with the environment.

Remarkably, once the fixed point of the TM is obtained, the excitation spectrum can be evaluated by constructing  one-dimensional excitation ans\"{a}tz ~\citep{2012_excitation}. 
The basic idea is to create an excitation by locally perturbing the fixed point and  make the momentum superposition to respect the translational symmetry [Fig.~\ref{fig:tangent_space}(f)]. 
The perturbed tenor $B$, once  restricting the excited state to be orthogonal to the fixed point, can then be found by maximizing the overlap of the expectation value of transfer matrix: $( \rho(B)|\mathbb{T}|\rho(B))$. 
However, if the fixed point is not unique, say, 2-fold degenerate, one should also consider the domain wall excitation which interpolates between two linearly independent fixed points [Fig.~\ref{fig:tangent_space}(f)]. 
In the case of the PEPS wave function considered in the main text, this corresponds to attaching a half-infinite $Z$ string on the right hand side of the perturbed $B$ tensor.

\section{Transfer Matrix Spectrum from Corner Transfer Matrix RG}
\label{sec:CTMRG}

\begin{figure}[b]
 \centering
\includegraphics[width=\linewidth]{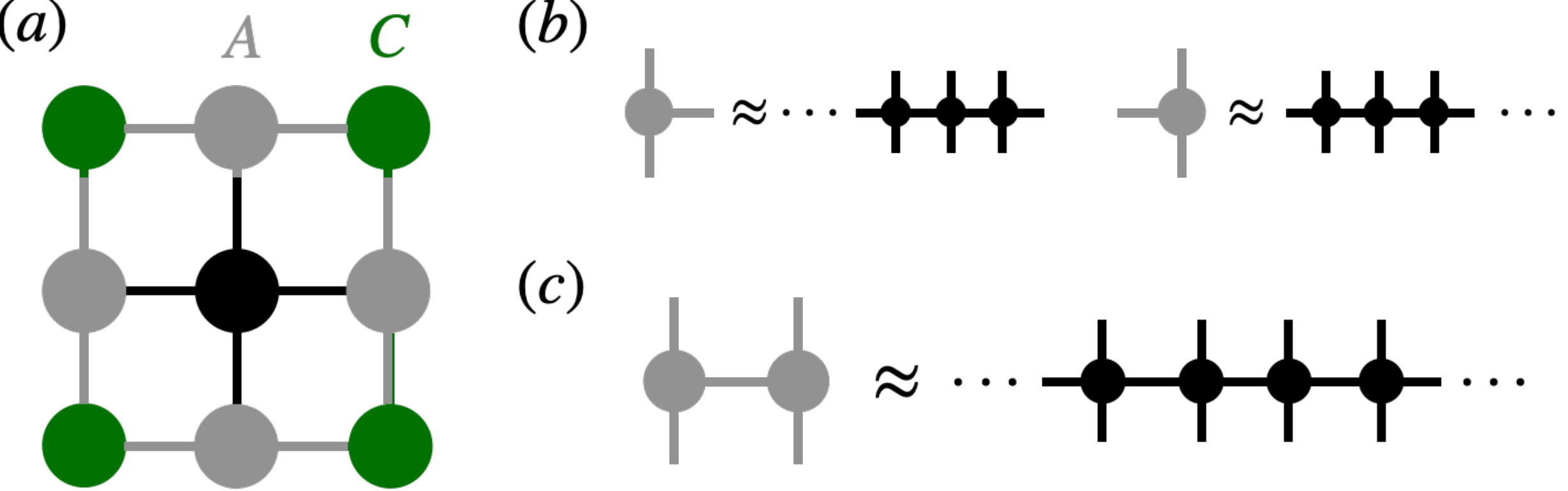}
\caption{(a) The environment of the CTMRG is represented by the corner transfer matrices $C$ and the boundary MPS $A$. (b) The left(right) boundary MPS is used to approximate the left(right)-infinite contraction of the double tensor. (c) The transfer matrix can be approximated by contracting the left and right boundary MPS.} 
\label{fig:CTM_scheme}
\end{figure}

\begin{figure}[t]
 \centering
\includegraphics[width=\linewidth]{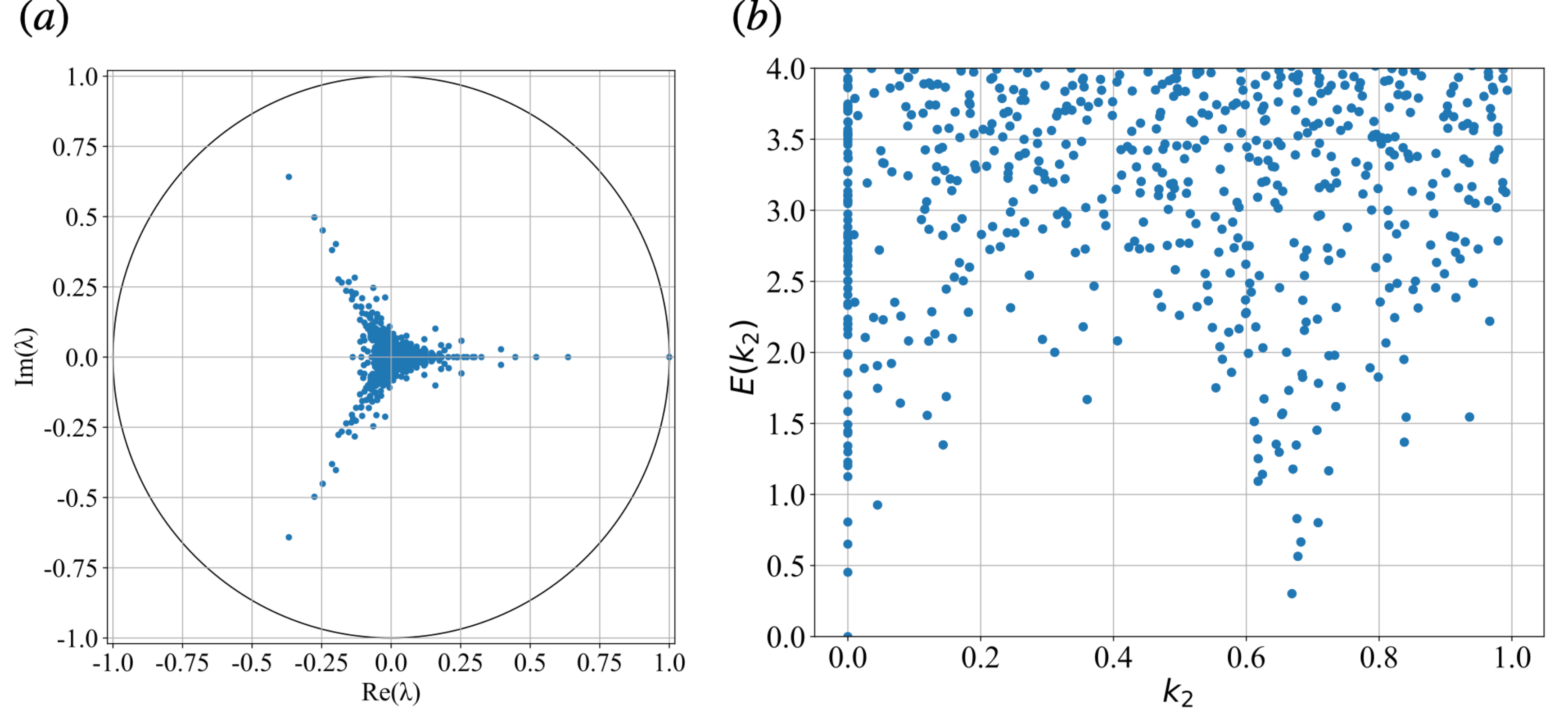}
\caption{(a) Eigenvalues $\lambda_j = e^{-E^j +ik_2^j}$ of  the TM from CTMRG on the complex plane within the unit circle. (b) Transfer matrix spectrum $E$ as a function of the phase $k_2$. } 
\label{fig:CTMRG_D8m60}
\end{figure}

The transfer matrix spectrum can also be obtained by a totally different way using CTMRG. In the CTMRG algorithm, the environment of the double tensor is represented by the corner transfer matrices $C$ and the boundary MPS $A$, which are  optimized iteratively [Fig.~\ref{fig:CTM_scheme}(a)]. Once the left(right) boundary MPS is obtained, it can be used to represent the left(right)-infinite contraction of the double tensor [Fig.~\ref{fig:CTM_scheme}(b)]. The transfer matrix can then be approximated by contracting the left and right boundary MPS [Fig.~\ref{fig:CTM_scheme}(c)]. Here the accuracy is also controlled by the dimension of boundary MPS $D_\text{mps}$. 
Once the approximate transfer matrix is obtained, the full spectrum can be evaluated using exact diagonalization. Thus, the accuracy is completely controlled by the MPS dimension $D_\text{mps}$ and no assumption of the single mode approximation is required.
In Fig.~\ref{fig:CTMRG_D8m60}(a), the  TM's eigenvalues using the same PEPS states as in the main text with $(D,D_\text{mps}) = (10, 60)$ are plotted on the complex plane within the unit circle. 
Most of the eigenvalues clearly arranged themselves along three distinct branches: $k_2 = 0,2\pi/3, -2\pi/3$. 
We observe the corresponding energies $E$ at $k_2 = \pm 2\pi/3$ are also lower than the energy at $\phi = 0$, which is consistent with the results obtained from the one-dimensional excitaiton ans\"{a}tz discussed in the main text. However, the CTMRG method does not give the momentum quantum number $k_1$ and is hard to distinguish the vacuum- and charge- sector excitations.

\bibliography{bibs}

\end{document}